\documentclass[twocolumn,showpacs,preprintnumbers,amsmath,amssymb,prb]{revtex4}
\usepackage{graphicx}% Include figure files
\usepackage{dcolumn}% Align table columns on decimal point
\usepackage{bm}% bold math

\begin{document}

\preprint{APS/123-QED}

\title{Ultrahigh Magnetic Field Phases in Frustrated Triangular-lattice Magnet CuCrO$_2$}% Force line breaks with \\

\author{Atsuhiko Miyata}
 \affiliation{Laboratoire National des Champs Magnetiques Intenses, CNRS-UGA-UPS-INSA, 143 Avenue de Rangueil, Toulouse 31400, France.}
 
 \author{Oliver Portugall}
 \affiliation{Laboratoire National des Champs Magnetiques Intenses, CNRS-UGA-UPS-INSA, 143 Avenue de Rangueil, Toulouse 31400, France.}
 
\author{Daisuke Nakamura}
 \affiliation{Institute for Solid State Physics, The University of Tokyo, 5-1-5 Kashiwanoha Kashiwa, Chiba 277-8581, Japan.}

\author{Kenya Ohgushi}\altaffiliation[Present address: ]{Department of Physics, Tohoku University, 980-8578 Miyagi, Japan.}
 \affiliation{Institute for Solid State Physics, The University of Tokyo, 5-1-5 Kashiwanoha Kashiwa, Chiba 277-8581, Japan.}

\author{Shojiro Takeyama}
\email{takeyama@issp.u-tokyo.ac.jp}
 \affiliation{Institute for Solid State Physics, The University of Tokyo, 5-1-5 Kashiwanoha Kashiwa, Chiba 277-8581, Japan.}

\date{\today}% It is always \today, today,

\begin{abstract}
The magnetic phases of a triangular-lattice antiferromagnet, CuCrO$_2$, were investigated in magnetic fields along to the $c$ axis, $H$ // [001],  up to 120 T . 
Faraday rotation and magneto-absorption spectroscopy were used to unveil the rich physics of magnetic phases. 
An up-up-down (UUD) magnetic structure phase was observed around 90--105 T at temperatures around 10 K. 
Additional distinct anomalies adjacent to the UUD phase were uncovered and the Y-shaped and the V-shaped phases are proposed to be viable candidates. 
These ordered phases are emerged as a result of the interplay of geometrical spin frustration, single ion anisotropy and thermal fluctuations in an environment of extremely high magnetic fields.

\end{abstract}

\pacs{75.30.Kz, 75.50.Ee, 78.20.Ls}

\maketitle

%\section{\label{sec:level1}First-level heading:\protect\\ The line
%break was forced \lowercase{via} \textbackslash\textbackslash}
 In geometrically frustrated magnets, the competition between different magnetic interactions produces highly degenerate magnetic ground states that are vulnerable to tiny perturbations, leading to diverse novel magnetic phases~\cite{textfrust}. 
 Among them, one typical state is a multiferroic state in which ferroelectricity is induced by unconventional magnetic structures that arise from geometrical magnetic frustration~\cite{kimura03, kimura06}. 
 Since changes in the spin structure alter the ferroelectricity, the application of magnetic fields plays an important role in elucidating the rich variety of magnetic and ferroelectric phases in geometrically frustrated magnet system. 
 
 Typical triangular-lattice antiferromagnets that are also multiferroic are CuFeO$_2$~\cite{kimura06} and CuCrO$_2$~\cite{seki08, kimura09}, both of which are delafossite oxides and have been intensively investigated in the past decade. 
 CuFeO$_2$ has a Curie-Weiss temperature of around -88 K and exhibits two successive phase transitions around 14 and 11 K~\cite{kimura06}. 
 Below 11 K, its magnetic structure becomes a four-sublattice collinear antiferromagnetic structure. When a magnetic field is applied to this state, a ferroelectric phase appears between $\sim$7 and 13 T, which is induced by a proper-screw magnetic structure. 
 This phenomenon is well described by a theoretical model proposed by Arima~\cite{arima07}. 
 
 Interestingly, additional magnetic phase transitions successively occur at higher magnetic fields in CuFeO$_2$, and magnetization plateaus with values of 1/5 and 1/3 of the saturation moment have actually been reported~\cite{lummen10}. 
 To date, some theoretical models were proposed to explain this rich occurrence of magnetic and ferroelectric phases. 
 For example, a theory proposed by Fishman~\textit{et al.}~\cite{fishman12} suggested a spin Hamiltonian into which magnetic interactions are incorporated up to the third-nearest neighbors as well as easy-axis single-ion anisotropy. 
 The importance of spin-phonon couplings was suggested by Wang and Vishwanath~\cite{wang08}. 
 However, none of these theories has been able to provide a general explanation of the magnetic and electric properties of multiferroic CuFeO$_2$ which therefore still remain an open issue. 
 
 To illuminate the complicated phases in delafossite oxides forming a triangular lattice, it is crucial to reveal the magnetic phases of another delafossite oxide, CuCrO$_2$, which has been known to have a much smaller easy-axis single-ion anisotropy $D$ with respect to its primary nearest-neighbor interaction $J_1$, in contrast to those of CuFeO$_2$~\cite{ye07, poienar10, yamaguchi10, fujita13}. For example, their ratio $D/J_1$ has been estimated by electron spin resonance (ESR) measuremetns as $D/J_1\sim0.017$ for CuCrO$_2$ and much smaller than $\sim0.097$ for CuFeO$_2$~\cite{yamaguchi10, fujita13}.
 
 CuCrO$_2$ has Curie-Weiss temperatures of -211 K (magnetic field applied perpendicular to the triangular-lattice plane) and -203 K (parallel to the plane), and exhibits two successive phase transitions around 24.2 and 23.6 K~\cite{seki08, kimura09}. 
 Below 23.6 K, its magnetic structure becomes an incommensurate proper-screw magnetic structure, as identified by neutron studies~\cite{soda09}, which induces ferroelectricity. 
 This mechanism is described by the theoretical model of Arima~\cite{arima07}.
 
 Remarkably, a recent study of CuCrO$_2$ under magnetic fields of up to 65 T applied parallel to the [001] axis by Mun~\textit{et al.} showed a rich magnetic-field-induced phase diagram including a few ferroelectric phases~\cite{mun14}, which are not reproduced by the theoretical model incorporating further-neighbor interactions and easy-axis single-ion anisotropy proposed for CuCrO$_2$ by Fishman~\cite{fishman11}. 
 Lin~\textit{et al.} conducted Monte Carlo calculations with a model including ÒspatiallyÓ anisotropic nearest-neighbor interactions and easy-axis single-ion anisotropy terms, which showed good agreement with their new results obtained from an experiment performed under higher magnetic fields up to 92 T~\cite{lin14}.
 
 As a consequence of their different magnetic interactions and anisotropy, both delafossite compounds, CuFeO$_2$ and CuCrO$_2$, show clearly different magnetic properties at low temperatures. Therefore, unveiling the high-magnetic-field phases of CuCrO$_2$ could provide further insight not only  into the rich magnetic and ferroelectric properties of this material but also of delafossite oxides in general.  
 
In this paper, we present magneto-optical studies (Faraday rotation and magneto-optical spectral absorption measurements) of CuCrO$_2$ carried out in ultrahigh magnetic fields up to 120 T and at temperatures down to 5 K. We reveal magnetic phases newly found in CuCrO$_2$, including the up-up-down (UUD) magnetic structure phase around 90--105 T at $\sim$10 K. 
 
 In our experiments a single-turn coil (STC) ultra-high magnetic field generator (UHMFG) at the Institute for Solid State Physics, University of Tokyo was used to generate magnetic fields exceeding 100 T~\cite{nakao85}. 
 Faraday rotation and magneto-optical spectral absorption measurements were conducted up to 120 T using a horizontally aligned STC-UHMFG. 
 The optical alignment around the STC was similar to that described in Ref. 19 and 20. Single crystals of CuCrO$_2$ were grown by a flux growth method using Bi$_2$O$_3$~\cite{kimura08}. 
 Plate-like samples parallel to the (001) crystal plane about 10 $\times$ 10 $\times$ 1 mm$^3$ in size were thus obtained.
  A sample of CuCrO$_2$ with 2 mm diameter was cut, then polished to 50 $\mu$m thickness and finally attached to a quartz substrate. 
  The magnetic field was applied parallel to the [001] axis in all measurements. A non-metallic helium-flow cryostat was used to cool down the sample to temperatures of $\sim$5 K~\cite{takeyama87}.
 
 Figure 1 shows the normalized magneto-optical transmission $T$($B$)/$T$(0) at a photon-energy of 1.943 eV (a wavelength of 638 nm), the Faraday rotation angle $\theta_\text{F}$, and the corresponding magnetization $M$ which is deduced by assuming their proportionality relation $\theta_\text{F} \propto M$, of CuCrO$_2$ under magnetic fields of up to 120 T at 5 K. 
 A magnetization curve obtained by Yamaguchi~\textit{et al.} using a non-destructing pulsed magnet up to 50 T at 1.3 K~\cite{yamaguchi10} is also shown by a dashed line as a reference in Fig. 1.  
 At 76 T, we observed clear anomaly associating a hysteresis in both magneto-optical transmissions $T$($B$)/$T$(0) and Faraday rotation angles $\theta_\text{F}$, indicating the first-order phase transition. 
 The anomaly was also observed in electric polarization measurements performed by Lin~\textit{et al.}~\cite{lin14} who suggested that it can be attributed to a phase transition from a commensurate Y-shaped phase (three spins form a ``Y'' shape) to the 1/3 magnetization plateau (UUD phase). 
 However, according to their Monte Carlo calculations a transition to the UUD phase cannot be of the first order. 
 In addition, the magnetic moment deduced from the FR angles turned out to be $\sim$0.83 $\mu$$_\text{B}$/Cr$^{3+}$ at 76 T, which is smaller than what would be expected for the 1/3 magnetization plateau (1 $\mu$$_\text{B}$/Cr$^{3+}$). 
 Therefore, it is natural to regard the phase just above 76 T as another magnetic phase prior to the UUD phase. 
 Details of this phase will be discussed later.

 The magnetization deduced from FR angles reaches 1 $\mu$$_\text{B}$/Cr$^{3+}$ around $\sim$95 T, but there is no clear evidence of a plateau-like phase in Fig.1. 
 The following scenario is the most likely: The 1/3 magnetization plateau is known to cause by an easy-axis anisotropy in classical Heisenberg triangular-lattice antiferromagnets~\cite{yun15}. 
 However, the anisotropy can be released by applying a magnetic field especially above the first-order phase transition at 76 T which is possibly associated with a lattice distortion. 
 Note that the easy-axis anisotropy of CuCrO$_2$ is rather small even in the absence of a magnetic field ($D/J_1\sim0.017$)~\cite{yamaguchi10}. 
 The reduction of the easy-axis anisotropy in magnetic fields has been taken into account, for example in the sister compound CuFeO$_2$, to explain its magnetic field induced phases~\cite{lummen10, fishman12}.
 
 Even without the easy-axis anisotropy, thermal fluctuations can induce the UUD phase in classical Heisenberg triangular-lattice antiferromagnets. This has been studied as the so-called ``order by disorder''~\cite{yun15, kawamura85}. However, the magnetization appears as an almost linear curve smeared out by temperature as shown in ref. 24. This is a viable reason why the 1/3 plateau is scarcely observed in current magnetization measurements.

\begin{figure}
   \centering
    \hfill
    \includegraphics[width=0.45\textwidth]{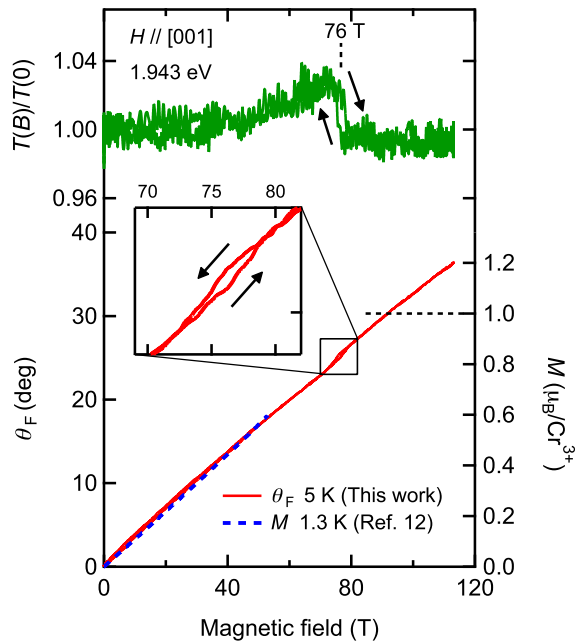}
    \hfill\null
\caption{\label{fig:Faradayrotation} 
(Color online) Normalized magneto-optical transmissions $T$($B$)/$T$(0) and Faraday rotation angles $\theta_\text{F}$ obtained by STC at 5 K together with magnetizations $M$ obtained at 1.3 K in ref. 12 ($H$ // [001]). Arrows show the hysteresis observed at 76 T.
}
\end{figure}

For further investigating details of the magnetic phases in CuCrO$_2$, we conducted magneto-optical transmission spectroscopy (MOTS) of excitonÐmagnon transitions (EMT). 
MOTS of EMTs is sensitive to magnetic phase transitions~\cite{miyata13, miyata11} because EMTs occur only when a magnon is required to compensate the spin and angular momentum changes of an otherwise optically forbidden excitonic transition. 
Spectral structures associated with EMT thus provide strong evidence for a change of both the magnetic and crystal structures. 
Fig. 1 demonstrates that $T$($B$)/$T$(0) responds in fact very sensitively to phase transitions.

Figure 2 shows the optical absorption (-log($T$($B$)) spectra of the $d$-$d$ transition and EMT in CuCrO$_2$ measured at 10 K; they are consistent with a previous report by Schmidt~\textit{et al.}~\cite{schmidt13}. 
The inset shows how the absorption spectrum evolves in magnetic fields up to 100 T in the wavelength region where EMT occurs. The peak intensity of the exciton-magnon absorption first decreases gradually up to 70 T and then increases with further increase of the magnetic field.

\begin{figure}
   \centering
    \hfill
    \includegraphics[width=0.45\textwidth]{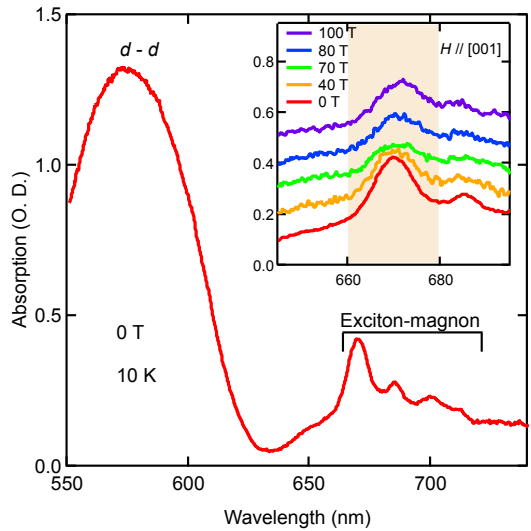}
    \hfill\null
\caption{\label{fig:spectra} 
(Color online) Optical absorption spectra of $d$-$d$ transition and EMT in CuCrO$_2$ at 0 T at 10 K. Inset shows magneto-optical absorption spectra in the region of EMTs at several magnetic fields along $H$ // [001] at 10 K obtained by streak spectroscopy. Absorption peak intensity between 660 and 680 nm is integrated and plotted as a function of magnetic field in Fig. 3.
}
\end{figure}

In Fig. 3 the intensity measured at wavelengths between 660 and 680 nm and at temperatures of 7 and 10 K is integrated (integrated absorption intensity: IAI) and plotted as a function of magnetic field together with the magnetization curve of CuCrO$_2$ deduced from the Faraday rotation angles at 5 K. 
Three distinct anomalies are observed at $\sim$75, 90, and 105 T in IAI (notified by black triangles in Fig. 3). 
The anomaly at $\sim$75 T corresponds to that observed at 76 T in the magnetization (the corresponding magnetic field differs slightly because of differences in the measurement temperature). 
A remarkable recovery of IAI is observed above $\sim$75 T. 
In conventional antiferromagnets, the EMT monotonically loses its intensity with increasing magnetic field, since a lesser number of magnons ($\Delta S_z=+1$) compensate for the spin angular momentum during exciton transition as the spin structure transforms to the canted configuration from the antiparallel configuration under magnetic fields. 
Magnon creation ($\Delta S_z=+1$) is quenched finally in a fully spin-polarized phase~\cite{eremenko75}. 

Therefore, the recovery of the EMT intensity reflects a change of the spin structure above $\sim$75 T. 
The Y-shaped spin structure is the most likely candidate. 
In this structure, spins approach an antiparallel configuration with increasing magnetic fields, which contributes to an increase of the excitonÐmagnon absorption intensity. 
In fact, the increase in the EMT intensity was observed in another multiferroic material, BiFeO$_3$, at the phase transition from the spin spiral to the canted antiferromagnetic phase, which causes an increase in ``antiparallelism''~\cite{xu09}. 

\begin{figure}
  \centering
    \hfill
    \includegraphics[width=0.45\textwidth]{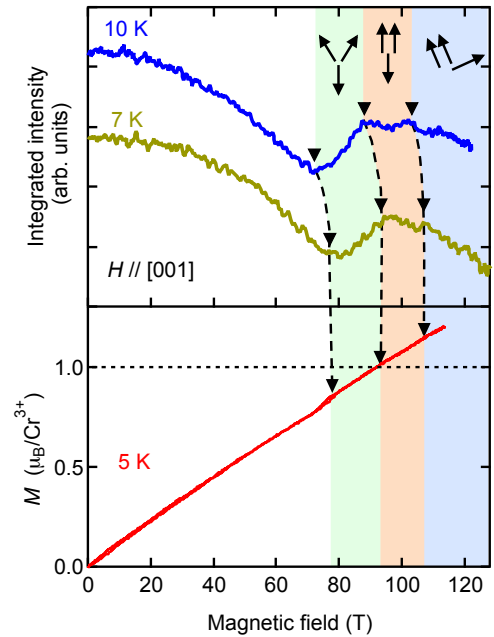}
    \hfill\null
\caption{\label{fig:EMT_int} 
(Color online) Integrated absorption intensity in the EMT of CuCrO$_2$ at 7 and 10 K at wavelengths between 660 and 680 nm (shaded area in inset of Fig. 2) and magnetization curve at 5 K deduced from Faraday rotation angles ($H$ // [001]). Arrows illustrate spin structures of Y-shaped, UUD, and V-shaped magnetic phases. Broken lines are a guide to the eyes for phase boundaries.
}

\end{figure}

In Fig. 3, around 90--105 T, the EMT intensity goes into a flat-top region (i.e.~the maximum of antiparallesim), which indicates that the spins form a collinear up-up-down structure (i.e.~UUD phase). The 1/3 plateau is scarcely visible in the magnetization ($M$) data. However, the deduced magnetization at 95 T reaches 1 $\mu$$_\text{B}$/Cr$^{3+}$, corresponding to the value expected for a 1/3 plateau. 
A slight widening of the flat-topped region is recognized upon increasing temperatures from 7 to 10 K. 
The UUD phase is known to stabilize by thermal fluctuations~\cite{yun15, kawamura85}. 
Above 105 T, the EMT intensity decreases again. 
A plausible magnetic phase above the UUD phase is a V-shaped magnetic phase in which two parallel spins and one other spin form a ``V'' shape (illustrated by arrows in Fig. 3). 
V-shaped and Y-shaped magnetic phases have been reported to respectively appear above and below the UUD phase in the phase diagram for a classical Heisenberg antiferromagnet on a triangular lattice with relatively weak easy-axis anisotropy~\cite{yun15}.

\begin{figure}
  
\end{figure}

\begin{figure}
   \centering
    \hfill
    \includegraphics[width=0.45\textwidth]{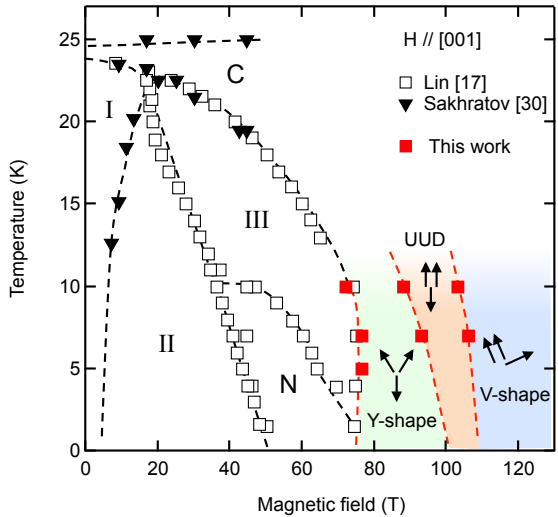}
    \hfill\null
\caption{\label{fig:phasediagram} 
(Color online) Magnetic phase diagram of CuCrO$_2$ ($H$ // [001]). For lower magnetic fields, the data are taken from Ref. 17 and 30. Arrows illustrate spin structures of Y-shaped, UUD, and V-shaped magnetic phases. Broken lines are a guide to the eyes for phase boundaries.
}
\end{figure}

Figure 4 shows the magnetic phase diagram of CuCrO$_2$ up to 120 T. 
The data for phase transitions in lower magnetic fields refer to measurements of the electric polarization $P$ ($H$ // [001] and $P$ // [110])~\cite{lin14} and nuclear magnetic resonance (NMR, $H$ // [001])~\cite{sakhratov16}. 
Sakhratov~\textit{et al.} have assigned the regions I and III to a three-dimensionally (3D) ordered incommensurate planar spin structure phase and a 2D-ordered (or 3D-polar) incommensurate planar spin structure phase, respectively~\cite{sakhratov16}. 
The region II is the intermediate phase of I and III with hysteretic behavior. 
At temperatures below 10 K the boundary of region ``N'' was observed in electric polarizations and assigned to an incommensurate umbrella-like spin structure (cycloidal spiral) phase~\cite{lin14}. 
The region ``C'' was attributed to a collinear spin-structure phase that could be connected to the collinear UUD phase that we observed around 90--105 T. 
The connection of two collinear phases has been theoretically suggested, since thermal fluctuations stabilize the UUD phase even at the zero limit of magnetic field~\cite{yun15}. 
This behavior has been observed in magnetic phase diagrams of other triangular-lattice Heisenberg antiferromagnets with easy-axis single-ion anisotropy, Rb$_4$Mn(MoO$_4$)$_3$~\cite{ishii11} and Ba$_3$MnNb$_2$O$_9$~\cite{lee14}. 

A striking difference between the magnetic phases of CuCrO$_2$ and CuFeO$_2$ is that collinear magnetic structures are unstable in CuCrO$_2$ in the low temperature limit. 
The collinear 1/5 magnetization plateau and collinear four-sublattice antiferromagnetic phase observed in CuFeO$_2$ have not been found in CuCrO$_2$. 
This difference arises from extremely small value of the easy-axis single-ion anisotropy of CuCrO$_2$ ($D/J_1\sim0.017$)~\cite{yamaguchi10} in contrast to that of CuFeO$_2$ ($D/J_1\sim0.097$)~\cite{fujita13}.

In summary, magneto-optical measurements of CuCrO$_2$ in ultrahigh magnetic fields up to 120 T applied along the [001] axis revealed that the UUD phase exists around 90--105 T at 7--10 K. 
Furthermore, additional anomalies were observed in the optical absorption intensities of the EMT, which revealed the existence of magnetic phases (presumably the Y-shaped and canted V-shaped phases) below and above the UUD phase. 
These magnetic phases emerge owing to the interplay of geometrical frustration, the magnetic field, and subtle perturbations of tiny easy-axis single-ion anisotropy and thermal fluctuations.

\section*{Acknowledgments}
We acknowledge Masayuki Hagiwara and Hironori Yamaguchi for sending their magnetization data which is shown in Fig. 1, and Kenta Kimura and Masashi Tokunaga for their helpful assistance to grow the samples. A. M.  thanks to a support of the Grant-in-Aid for the Japan Society for the Promotion of Science (JSPS) Fellows.

\end{document}